\documentclass{emulateapj}

\shorttitle{VLBI cores in Seyferts}\shortauthors{Giroletti \& Panessa}

\begin{document}

\title{The faintest Seyfert radio cores revealed by VLBI}

\author{Marcello Giroletti\altaffilmark{1} and Francesca Panessa\altaffilmark{2}}

\email{giroletti@ira.inaf.it, francesca.panessa@iasf-roma.inaf.it}

\altaffiltext{1}{INAF Istituto di Radioastronomia, via Gobetti 101, 40129
  Bologna, Italy}

\altaffiltext{2}{IASF/INAF, via del Fosso del Cavaliere 100, 00133 Roma, Italy}

\defcitealias{Ho2001}{HU01}
\defcitealias{Perez-Torres2007}{P{\'e}rez-Torres \& Alberdi 2007}

\begin{abstract}

In this letter, we report on dual-frequency European VLBI Network (EVN)
observations of the faintest and least luminous radio cores in Seyfert nuclei,
going to sub-mJy flux densities and radio luminosities around
$10^{19}\,$W\,Hz$^{-1}$.  We detect radio emission from the nuclear region of
four galaxies (NGC\,4051, NGC\,4388, NGC\,4501, and NGC\,5033), while one
(NGC\,5273) is undetected at the level of $\sim 100\, \mu$Jy. The detected
compact nuclei have rather different radio properties: spectral indices range
from steep ($\alpha>0.7$) to slightly inverted ($\alpha = -0.1$), brightness
temperatures vary from $T_B=10^5$ K to larger than $10^7$ K and cores are
either extended or unresolved, in one case accompanied by lobe-like features
(NGC\,4051). In this sense, diverse underlying physical mechanisms can be at
work in these objects: jet-base or outflow solutions are the most natural
explanations in several cases; in the case of the undetected NGC\,5273 nucleus,
the presence of an advection-dominated accretion flow (ADAF) is consistent with
the radio luminosity upper limit.

\end{abstract}


\section{Introduction}

Active Galactic Nuclei (AGN) are traditionally divided in radio quiet (RQ) and
radio loud (RL). The latter are typically powerful radio sources
($P_r>10^{22}\,$W\,Hz$^{-1}$), with large scale radio lobes and bright compact
cores; VLBI observations routinely target their nuclear regions, showing high
brightness temperatures and in some cases jet knots with superluminal
motions. Radio quiet AGN, such as Seyfert galaxies, are much fainter radio
sources and their radio emission is confined to the sub-kpc scale. However,
moderately deep VLA surveys show that most AGN are radio sources at some level
\citep[][~hereinafter HU01]{Nagar2002,Ho2001}.  While the origin of the radio
emission in RL AGN is well established as synchrotron radiation from energetic
particles in jets and lobes, the case for RQ AGN is much less clear.

Since nuclear structures in most Seyfert galaxies show complex morphologies, it
is of fundamental importance to resolve them with high spatial
resolution. Indeed, it has been shown that VLBI observations of the pc-scale
and subpc-scale region of Seyfert nuclei and Low-luminosity AGNs (LLAGNs) are
often successful in the determination of the physical parameters of the nuclear
radio components (such as the brightness temperature, the spectral index, the
jet motions, etc.) and therefore in the comprehension of the underlying
physical mechanisms. For example, the VLBA study of the classical Seyfert 2
galaxy NGC\,1068 has allowed the identification of the location of the hidden
active nucleus and the attribution of the core radio emission to thermal
free-free emission from an X-ray heated corona or wind arising from the disk
\citep{Gallimore2004}. In the case of the type 1 Seyfert NGC~4151 it has been
possible to resolve the 0.2 pc two-sided base of a jet whose low speeds
indicate non-relativistic jet motions, possibly due to thermal plasma
\citep{Ulvestad2005}. On the contrary, the VLBA analysis of the LLAGN NGC\,4278
has shown that the radio emission of this source is emitted via synchrotron
process by relativistic particles similarly to ordinary radio-loud AGN
\citep{Giroletti2005}. These studies indicate that the analysis of individual
sources is very important for the determination of the different spatial
components and physical parameters.

The five sources here presented belong to a complete distance limited ($d <
22\,$Mpc) sample of 27 nearby Seyfert galaxies \citep{Cappi2006}. Accurate
multi-wavelength studies are available for this sample. In particular, VLA
radio images at 1.4 and 5 GHz are presented in \citetalias{Ho2001}, and VLA
data at 15 GHz are also available for $25$ sample sources
\citep{Nagar2002}. VLBI observations have become available for a few sources
through the years; $9$ sources of the sample have 5 GHz VLBA observations
\citep{Nagar2002}, while for the radio brightest galaxies, e.g.\ NGC\,1068 and
NGC\,3031, dedicated works are available \citep{Gallimore2004,Bietenholz2004}.
However, most of the weakest sources in the sample have never been observed
with milliarcsecond resolution.

In the present letter, we report on new high sensitivity VLBI observations of
five sources with $S\sim1$ mJy, aimed at answering two fundamental questions:
(1) how common/frequent is the presence of (sub-)parsec scale radio sources
even in the weakest nuclei and (2) what do the properties of the detected radio
sources tell us about the physics of individual sources and the viability of
jet-base versus Advection-Dominated Accretion Flow \citep[ADAF,][]{Narayan1994}
explanations. The observations are detailed in \S2 and the results presented in
\S3. A discussion and the main conclusions are given in \S4 and \S5,
respectively. Throughout the paper, we define spectral indices such that
$S(\nu)\propto \nu^{-\alpha}$ and adopt $H_0=70$ km s$^{-1}$ Mpc$^{-1}$,
$\Omega_\Lambda=0.73$ and $\Omega_m=0.27$ \citep{Spergel2003}.

\section{Observations}

We observed the nuclei of five Seyfert galaxies with the European VLBI Network
(EVN) at 1.6 and 5 GHz, namely \object{NGC\,4051}, \object{NGC\,4388},
\object{NGC\,4501}, \object{NGC\,5194}, and \object{NGC\,5273}; a basic summary
of the observational parameters is given in Table~\ref{t.log}.  We used the
maximum recording rate of 1 Gbps and each source was observed for about 6 hrs,
switching between targets in order to improve the coverage of the
$(u,v)$-plane. Amplitude calibration was done {\it a priori} in the standard
EVN pipeline; phase corrections were obtained using observations of bright
nearby ($0.2^\circ<d<3.2^\circ$) calibrators, in the so-called phase
referencing technique.  The same calibrators were used to check the consistency
of the amplitude calibration, which results accurate to within 10\%.

Phase coherence was clearly detected on the baselines with large sensitive
telescopes in four cases at 1.6 GHz and in two cases at 5 GHz. Significant
weather and hardware problems affected some of the largest apertures during the
5 GHz observations, resulting in generally higher noise levels and lack of
short spacings, which is a serious problem given the characteristics of our
targets. The typical off-source noise level is $20-40\, \mu$Jy beam$^{-1}$ at
1.6 GHz, and $60-100\, \mu$Jy beam$^{-1}$ at 5 GHz. Owing to the low flux
density of the targets, no self-calibration was attempted.

We derive flux densities and positions from Gaussian fits to the image plane in
AIPS. As a check, we have also performed a visibility model fit in Difmap in
the cases where the signal-to-noise ratio is sufficient to well constrain the
best fit parameters. The results turn out to be in agreement. The final
parameters are reported in Table \ref{t.mf}, along with the resulting spectral
indices and brightness temperature limits. Positions are generally determined
with sub-mas accuracy ($\Delta r \sim 0.5$ mas).

%, while the uncertainty on the
%flux density estimate is around 10\% (mostly dependent on the absolute {\it a
%  priori} amplitude calibration process).

\section{Results}

Radio emission is detected at both 1.6 and 5 GHz in the cores of NGC\,4051 and
NGC\,5033, at 1.6 GHz with upper limits at 5 GHz in NGC\,4388 and NGC\,4501,
while NGC\,5273 remains undetected at either frequencies. The detected galaxies
have a flux density at milliarcsecond scales that ranges from little more than
1 mJy down to about the threshold set by sensitivity limit of our observation
($\sim$ a few 100's $\mu$Jy). In general, this corresponds to a fraction
between 5 and 40\% of the corresponding VLA flux density in the same sources.

The estimated sizes are of order $\sim10$ mas (FWHM), although several
components are unresolved and could actually be more compact. At the distance
of our targets ($D<19$ Mpc), this corresponds to linear sizes $<0.9$
parsec. The brightness temperatures limits for such components have been
calculated assuming sizes of $1/2$ of the model-fit FWHM, and are typically in
excess of $T_B>10^6$ K. Details on the single sources are given in the
following subsections.

\subsection{NGC\,4051}

NGC\,4051 is a Seyfert 1.2 galaxy with a complex radio structure at all
scales. On scales of a few arcseconds, it is dominated by emission extending
toward the southwest \citepalias{Ho2001}, while at higher resolution the
central core splits into a small-scale double/triple separated by $0.4\arcsec$,
roughly in the east-west direction \citep{Ulvestad1984,Kukula1995}.

\begin{figure*}
\plotone{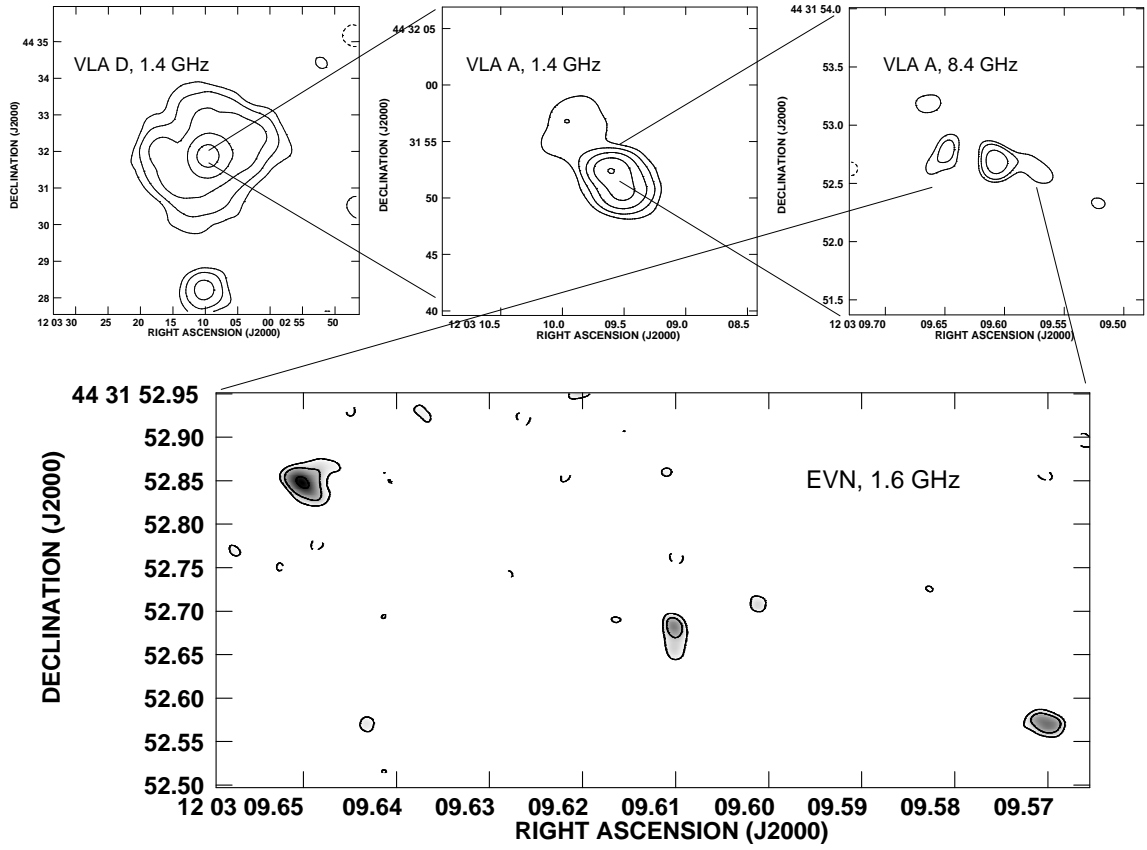}
\caption{NGC\,4051. Top panels are VLA data at increasing resolution from a
  reanalysis of archival data taken on March 1995 (left), August 1999 (middle),
  and June 1991 (right); lowest contour and peak brightness are (1.1, 23.5),
  (0.4, 6.5), and (0.2, 0.7) mJy beam$^{-1}$, respectively. Bottom panel: our
  EVN image at 1.6 GHz, with contours traced at $(-3, 3, 5, 10) \times 34\,
  \mu$Jy beam$^{-1}$. \label{f.4051}}
\end{figure*}

We detect radio emission at sub-parsec scales in NGC\,4051 at both 1.6 and 5
GHz. At 1.6 GHz, we detect three sub-mJy components (see
Fig.\ \ref{f.4051}). Two of these, separated by 20 parsecs, are clearly
associated with the small-scale double structure imaged with the VLA by
\citet{Ulvestad1984}.  The third component (0.67 mJy) is symmetric to the
easternmost one with respect to the central one, and aligned with the
southwestern emission revealed at larger scale. This is also associated with
emission at 8.4 GHz as shown by \citet{Kukula1995} and in the top right panel
of Fig.\ \ref{f.4051}.

The central component is detected also at 5 GHz, with a moderately steep
spectral index ($\alpha=0.7$). The upper limit for the source radius is 0.31 pc,
which corresponds to $\sim 2.5 \times 10^6 R_S$, a factor of $\sim$ 50 larger
than the measured BLR size \citep[$\sim$ 0.006 pc,][]{Kaspi2000}.

\subsection{NGC\,4388}

Arcsecond and sub-arcsecond resolution images of this Sy\,1.9 galaxy have
revealed a complex morphology with a compact, flat spectrum component detected
up to 15 GHz \citep{Carral1990,Falcke1998}, while the only previous VLBI
observation of this galaxy resulted in a non detection with the VLBA at 8.4 GHz
\citep{Mundell2000}.

We clearly detect the source at 1.6 GHz, with a flux density of 1.3 mJy and an
extension of 0.48 pc ($\sim 6$ mas). This is in agreement with the lower and
upper limits, derived by \citet{Mundell2000} and \citet{Carral1990},
respectively, which constrained the emitting region to a size $3<
\theta/\mathrm{mas} < 70$.

The 5 GHz data are difficult to interpret owing to the almost complete lack of
data from the Effelsberg-Westerbork and Effelsberg-Jodrell baselines. Since
these are the shortest and most sensitive baselines in our array, it is hard to
tell whether the non detection has to be ascribed to the large rms noise, the
too narrow resolution, or a combination of both. Still, the result remains
somewhat surprising in the light of the flat spectral index observed with the
VLA.

\subsection{NGC\,4501}

Known also as M\,88, NGC\,4501 is a Sy\,1.9 galaxy hosting an unresolved VLA
radio source \citepalias{Ho2001}. We detected this source at 1.6 GHz, as an
unresolved component with a peak brightness of 0.71 mJy\,beam$^{-1}$. This
corresponds to a brightness temperature in excess of $5.5 \times 10^6$ K and a
linear size of 0.6 pc (or $8 \times 10^4R_S$).

Our data thus account for a 40\% fraction of the VLA detected flux density at
1.4 GHz, suggesting that the radio source is quite compact. The data for this
source are subject to the same observational limits reported for NGC\,4388, so
that it is difficult to tell if the non detection is due to a moderately steep
spectral index, or a resolved structure at higher frequency.

\subsection{NGC\,5033}

VLA data at 1.4 and 5 GHz show a slightly resolved core in this Sy\,1.5 galaxy,
with a short jet-like extension to the East
\citepalias{Ho2001,Perez-Torres2007}. We detect this source at both 1.6 and 5
GHz, revealing an unresolved, flat spectrum component. Indeed, this source
shows the flattest spectral index ($\alpha_{1.6}^5=-0.1\pm0.1$), the highest
brightness temperature limit ($T_B>1.3\times10^7$ K), and the smallest size
limit ($d<0.27$ pc, equivalent to $1.4\times 10^5 R_S$) among our sample.

\subsection{NGC\,5273}

VLA observations of the Sy\,1.5 galaxy NGC\,5273 had revealed an unresolved
component ($\theta<0.52\arcsec$) at 1.4 and 5 GHz, with a rather flat spectral
index \citepalias[$\alpha_{1.4}^5=0.4$,][]{Ho2001}.  \citet{Nagar1999} had also
revealed a 0.6 mJy compact component ($\theta<0.3\arcsec$) at 8.4 GHz.

It is therefore remarkable that our observations do not reveal any emission
from the nuclear region of this source at neither 1.6 or 5 GHz. The $3\sigma$
upper limit peak brightness is $90\, \mu$Jy beam$^{-1}$ at 1.6 GHz and $140 \,
\mu$Jy at 5 GHz. This implies either a significant resolution of the
sub-arcsecond emission ($>95\%$ of the flux density at angular scales
$20<\theta/\mathrm{mas}<300$), significant variability (by $\sim$ one order of
magnitude), or a combination of both.

\section{Discussion and conclusions}

With the VLBI observations presented in this letter, we have targeted the
faintest and least luminous nuclei among well known local AGN, going to sub-mJy
flux densities and radio luminosities around $10^{19}$ W Hz$^{-1}$.

These sources belong to a complete sample of nearby Seyfert galaxies, in which
previous VLBI observations -- albeit limited to the brightest members -- had
revealed an ubiquitous presence of sub-parsec cores and/or structures.  Similar
findings had been reported on other relatively bright flat spectrum radio
sources in LLAGN; for instance, \citet{Nagar2002} found that almost all LLAGNs
in the Palomar sample with S$_\mathrm{VLA, 15\, GHz}> 2.7$ mJy show mas-scale
or sub mas-scale radio cores.  The 80\% detection rate in the present
sub-sample extends these findings to the lowest luminosity and flux density
regimes.

While the presence of sub-pc scale radio emission appears thus to be
ubiquitous, it is remarkable that it accounts only for a fraction between 5\%
and 40\% of the emission detected on scales of a few tens of parsecs (eg.\ as
seen in VLA observations). As a consequence, a large fraction of the parsec
scale radio luminosity is emitted in a diffuse region.  This could be the case
of NGC\,4388 and NGC\,4501, in which we completely resolve the VLA sources at 5
GHz, but detect them at 1.6 GHz (size $\ga 10$ mas). Similarly, the non
detection of NGC\,5273 requires that $>95\%$ of the VLA 1.6 GHz flux density is
emitted on scales larger than 20 mas (1.6 pc).

Even though much of the intermediate scale emission is resolved out, the
majority of the nuclear regions do host compact radio sources at sub-parsec
scales. In Table~\ref{tabtot}, we report the multi-wavelength properties of the
sources presented in this letter.  The common features in these nuclei are: (i)
their extremely low radio luminosity (Cols.\ 4-6), at the level of the least
luminous Seyfert nuclei \citep[such as the sub-mJy source
  NGC\,4395,][]{Wrobel2006}; (ii) their extreme radio-quietness, for instance
the ratios between the radio and nuclear X-ray emission (Col.\ 8) are among the
lowest ever measured for LLAGN \citep{Panessa2007,Terashima2003}; and (iii)
their low Eddington ratios (Col.\ 9), a common trait in low luminosity nuclei
\citep{Ho2009}.

Radio emission from LLAGN cores is generally associated with accretion/ejection
processes in the vicinity of a supermassive black hole
\citep{Falcke2000,Ulvestad2001,Nagar2002}. While some of the resolved extended
emission could be of thermal origin, the small linear scales ($< 0.6$ pc) and
comparatively high brightness temperatures ($T_B=10^{5-7}$ K) measured in our
sources suggests that the weakest Seyfert nuclei could also be scaled-down
versions of more luminous AGN. In this scenario, the origin of radio emission
can be generally attributed to synchrotron emission from the base of a jet
coupled with a low-power accretion disk \citep{Falcke1999}.  Alternatively, the
presence of an ADAF \citep{Narayan1994} is invoked to explain the low
luminosity of these sources, although the predicted radio emission often fails
to reproduce the data, requiring combined jet/ADAF models \citep[see
  e.g.,][]{Yuan2002}. On average, our results do not seem consistent with the
presence of an ADAF alone, on the basis of the observed structures, sizes, flux
densities, and spectral indices.

The two sources detected at both 1.6 and 5 GHz (NGC\,4051 and NGC\,5033) are
most easily explained in terms of jet-base/outflow phenomena. First, the
detection of three aligned sub-mJy components in NGC\,4051 is indeed suggestive
of ejection processes. The brightness temperature ($T = 2\times10^5$ K) does
not completely rule out a thermal origin, consistent with the presence of an
outflow rather than a relativistic jet, similar to the case of NGC\,4395, which
shows an elongated low brightness temperature nuclear structure, tracing
possible outflow emission \citep{Wrobel2006,Christopoulou1997}.  The EVN radio
core is also positionally coincident with a low luminosity H$_2$O maser,
suggesting that the radio continuum may arise from the inner regions of a
molecular disk or from a nuclear wind \citep{Hagiwara2007}.

As for NGC\,5033, its core presents all the hallmarks of a jet-base feature,
being detected and unresolved at both at 1.6 and 5 GHz, with a brightness
temperature lower limit of $T_B>1.3\times10^7$ K, and a flat spectrum
(consistent with $\alpha=0$).  While these would somewhat be consistent with
ADAF expectations, the 5 GHz luminosity is $\sim$ 4 times in excess of what
predicted on the basis of the observed L$_\mathrm{2-10 keV}$ luminosity,
implying that the ADAF model alone fails to account for the observed radio
emission. The presence of a jet or an outflow component is therefore required
and more consistent with the data. Indeed, the short jet-like extension found
at larger scale with VLA \citepalias{Ho2001,Perez-Torres2007} favours the
jet-base hypothesis.

With all the caveats related to the unfavourable observing conditions at 5 GHz,
the two sources detected only at 1.6 GHz (NGC\,4388 and NGC\,4501) seem to be
also at odds with an ADAF scenario. The non detection at high frequency points
to a steep spectral index and/or to a rather extended structure, at scales of
$10^6 R_S$ or more, in contrast with the flat/inverted spectrum and the compact
structure predicted by ADAF models. Although consistent with an angular scale
of several mas, thermal emission is also unlikely given the $\ga 10^6$ K
brightness temperatures observed at 1.6 GHz. NGC\,4388 and NGC\,4501 are also
the only type 1.9 Seyferts in our sample, showing heavily absorbed/weak X-ray
spectra \citep{Cappi2006}.

Interestingly, the only undetected source in our sample is NGC\,5273, a type
1.5 Seyfert, in which the nucleus is seen directly, displaying broad emission
lines and a bright X-ray spectrum \citep{Cappi2006}. Either the VLA emission is
resolved out or the source is variable. Indeed, the source was initially not
revealed at 8 GHz with the VLA \citep[$S < 0.23$ mJy,][]{Kukula1995} and lately
detected with $S = 0.6$ mJy by \citet{Nagar1999}. The comparison between the
X-ray and the EVN radio luminosity upper limit reveals that this source is
extremely radio quiet. Indeed, this is the only case in which the radio data
are consistent with a pure ADAF accretion mechanism, since our upper limit for
this nucleus is 4 times above the radio core luminosity derived from the
observed X-rays \citep{Yi1998}.

\section{Conclusions}

The five Seyfert galaxies presented in this letter are the faintest observed to
date among a larger sample, extending the availability of VLBI observations to
the regime of $\la 1 $ mJy flux density. With compact components detected in
4/5 of our sources, we confirm that parsec scales radio emission is almost
ubiquitous in Seyfert nuclei.

Despite this common trait, the observational and derived physical parameters
display heterogeneous behaviours, not easy to interpret within a common
physical scenario.  Spectral indices range from quite steep ($\alpha>0.7$) to
slightly inverted ($\alpha = -0.1$) and brightness temperatures between
$T_B=10^5$ K and $>10^7$ K; the emitting regions are either extended or
unresolved, in one source accompanied by lobe-like features and in an extreme
case not even detected.

As long as the present data set seems generally in favour of jet-like or
outflow interpretation for most sources, more data are necessary for a complete
understanding of the physical processes at work. The peculiar double/triple
morphology of NGC\,4051 is certainly worth a dedicated project (e.g.\ to look
for proper motion of the external components); it also seems necessary to
constrain the angular size and spectral index of NGC\,4388 and NGC\,4501 with
new 5 GHz data, not affected by the severe failures on short baselines faced by
the present observations. Finally, even deeper observations are needed to
constrain the viability of an ADAF scenario in NGC\,5273, as well as to probe
the faintest members of the parent sample.

\acknowledgments The European VLBI Network is a joint facility of European,
Chinese, South African and other radio astronomy institutes funded by their
national research councils.  This work has benefited from research funding from
the European Community's sixth Framework Programme under RadioNet R113CT 2003
50 58187.

{\it Facilities:} \facility{EVN}, \facility{VLA}.

\newpage 

\begin{deluxetable*}{llclc} % for use with referee option
\tabletypesize{\small}                     % for use with referee option
%\rotate
\tablecaption{Log of observations \label{t.log}}
\tablewidth{0pt}
\tablehead{
\colhead{} & \multicolumn{2}{c}{1.6 GHz observations} & \multicolumn{2}{c}{5 GHz observations} \\
\colhead{Galaxy} & \colhead{Date} & \colhead{FWHM} & \colhead{Date} & \colhead{FWHM} \\
\colhead{} & \colhead{} & \colhead{(mas $\times$ mas, $^\circ$)} & \colhead{} & \colhead{(mas $\times$ mas, $^\circ$)} \\
\colhead{(1)} &\colhead{(2)} &\colhead{(3)} &\colhead{(4)} &\colhead{(5)} 
}
\startdata
NGC\,4051 & 2007 Jun 6  & $13.5 \times 11.3, 19$ & 2007 Jun 1 & $9.0 \times 6.4, 50$  \\
NGC\,4388 & 2008 Feb 28 & $26.1 \times 20.9, 33$ & 2008 Mar 10 & $4.2 \times 1.3, 6$ \\
NGC\,4501 & 2008 Feb 28 & $13.8 \times 9.6, 2$   & 2008 Mar 10 & $4.1 \times 1.3, 6$ \\
NGC\,5033 & 2007 Jun 6  & $11.1 \times 6.2, -26$ & 2007 Jun 1 & $5.0 \times 3.5, 76$ \\
NGC\,5273 & 2007 Jun 6  & $11.5 \times 7.1, -24$ & 2007 Jun 1 & $5.4 \times 4.1, 70$ \\
\enddata
\tablecomments{Col.\ (1): galaxy name. Col.\ (2): date of 1.6 GHz observation. Col.\ (3): beam size and orientation at 1.6 GHz. Col.\ (4): date of 5 GHz observation. Col.\ (5): beam size and orientation at 5 GHz.}
\end{deluxetable*}          % for use with referee option

\begin{deluxetable*}{ccrrrrrrrrr} % for use with referee option
\tabletypesize{\footnotesize}                     % for use with referee option
\tablecaption{EVN observational results and model fit parameters \label{t.mf}}
\tablewidth{0pt}
\tablehead{
\colhead{Galaxy} & \colhead{RA (J2000)} & \colhead{Dec. (J2000)} & \colhead{Size$_{1.6}$, PA$_{1.6}$} & \colhead{$S_{1.6}$} & \colhead{Size$_{5}$, PA$_{5}$} & \colhead{$S_5$} & \colhead{$\alpha_{1.6}^5$} & \colhead{$T_B$} \\
\colhead{} & \colhead{$(h\, m\, s)$} & \colhead{$(^\circ \, \arcmin \, \arcsec)$} & \colhead{(mas $\times$ mas, $^\circ$)} & 
\colhead{(mJy)} & \colhead{(mas $\times$ mas, $^\circ$)} &  \colhead{(mJy)} & \colhead{} & \colhead{(K)} \\
\colhead{(1)} &\colhead{(2)} &\colhead{(3)} &\colhead{(4)} &\colhead{(5)} &\colhead{(6)} &\colhead{(7)} &\colhead{(8)} &\colhead{(9)} 
}
\startdata
NGC\,4051 & 12 03 09.6102 & +44 31 52.678 & $24.8 \times 8.8, 26^\circ$ & $0.45\pm0.05$ & $<8.2 \times 7.5, 61^\circ$ & $0.20\pm0.02$ & $0.7 \pm 0.1$ & $>1.8 \times 10^5$  \\
\nodata   & 12 03 09.6504 & +44 31 52.848 & $24.3 \times 15.4, 71^\circ$ & $0.73\pm0.07$ & \nodata & \nodata & \\
\nodata   & 12 03 09.5700 & +44 31 52.572 & $29.1 \times 10.3, 81^\circ$ & $0.67\pm0.07$ & \nodata & \nodata & \\
NGC\,4388 & 12 25 46.7814 & +12 39 43.768 & $15.0 \times 5.9, 12^\circ$ & $1.32\pm0.13$ & \nodata & $<0.55$\tablenotemark{a} & $>0.7$ & $1.3 \times 10^6$ \\
NGC\,4501 & 12 31 59.1529 & +14 25 13.169 & $<11.2 \times 7.5, -5^\circ$ & $0.73\pm0.07$ & \nodata & $<0.33$\tablenotemark{a} & $>0.6$ & $>5.5 \times 10^6$ \\
NGC\,5033 & 13 13 27.4711 & +36 35 37.924 & $<5.6 \times 5.8, 19^\circ$ & $0.68\pm0.07$ & $<3.0 \times 3.0, -42^\circ$ & $0.76\pm0.08$ & $-0.1 \pm 0.1$ & $>1.3 \times 10^7$ \\
\enddata
\tablenotetext{a}{The 5 GHz upper limits for NGC\,4388 and NGC\,4501 are $3\sigma$ rms noise levels measured on images with the same resolution of the 1.6 GHz ones}
\tablecomments{Col.\ (1): galaxy name. Col.\ (2, 3): absolute coordinates of radio position. Col.\ (4): size and position angle (PA) at 1.6 GHz. Col.\ (5): flux density at 1.6 GHz. Col.\ (6): size and PA at 5 GHz. Col.\ (7): flux density at 5 GHz. Col.\ (8): spectral index between 1.6 and 5 GHz. Col.\ (9):  brightness temperature.}
\end{deluxetable*}          % for use with referee option

\begin{deluxetable*}{lrrrrrrrrrr}          % for use with referee option
\tabletypesize{\footnotesize}                     % for use with referee option\tablecaption{Properties of the Seyfert galaxies}
\tablecaption{Properties of the Seyfert galaxies}
\tablewidth{0pt}
\tablehead{
\colhead{Galaxy} & \colhead{D} & \colhead{Class} & \colhead{Log P$_{1.6}$} &
\colhead{Log L$_{1.6}$} & \colhead{Log L$_{5}$} & \colhead{Log L$_\mathrm{X}$} &
\colhead{Log L$_{5}$/L$_\mathrm{X}$} & \colhead{Log L$_\mathrm{X}$/L$_\mathrm{Edd}$} & 
\colhead{Log M$_\mathrm{BH}$} &\colhead{Log $R_\mathrm{S}$}  \\
\colhead{} & \colhead{(Mpc)} &\colhead{} & \colhead{(W Hz$^{-1}$)} & \colhead{(ergs s$^{-1}$)} & \colhead{(ergs s$^{-1}$)} & \colhead{(ergs s$^{-1}$)} &\colhead{} & \colhead{} & \colhead{($M_\odot$)} & \colhead{(pc)}  \\
\colhead{(1)} &\colhead{(2)} &\colhead{(3)} &\colhead{(4)} &\colhead{(5)} &\colhead{(6)} &\colhead{(7)} &\colhead{(8)} &\colhead{(9)} &\colhead{(10)} &\colhead{(11)} 
}
\startdata
NGC\,4051 &  9.3 & S1.2 & 18.7 & 34.9 & 35.0 & 40.8 & $-$5.80 & $-$3.41 & 6.11 & $-$6.89  \\
NGC\,4388 & 16.7 & S1.9 & 19.6 & 35.8 & $<$35.7 & 41.7 & $<-$6.06 & $-$3.17 & 6.80 & $-$6.20  \\
NGC\,4501 & 16.8 & S1.9 & 19.4 & 35.6 & $<$35.4 & 39.6 & $<-$4.21 & $-$6.39 & 7.90 & $-$5.10  \\
NGC\,5033 & 18.7 & S1.5 & 19.4 & 35.7 & 36.2 & 41.1 & $-$4.90 & $-$4.31 & 7.30 & $-$5.70  \\
NGC\,5273 & 16.5 & S1.5 & $<$18.5 & $<$34.7 & $<$35.4 & 41.4 & $<-$6.02 & $-$3.24 & 6.51 & $-$6.49  \\
\enddata
\tablecomments{Col.\ (1): galaxy name. Col.\ (2): galaxy distance. Col.\ (3):  optical classification. Col.\ (4): monochromatic power at 1.6 GHz. Col.\ (5): luminosity at 1.6 GHz. Col.\ (6): luminosity at 5 GHz. Col.\ (7): X-ray 2-10 keV luminosity. Col.\ (8): ratio of radio to X-ray luminosity. Col.\ (9): ratio of X-ray to Eddington luminosity. Col.\ (10): black hole mass. Col.\ (11): Schwarzschild radius.  Data in cols.\ (2), (3), (7) and (10) are taken from \citet{Panessa2007}, except for NGC\,4051, whose distance is from \citet{Barbosa2006}. }
\label{tabtot}
\end{deluxetable*}          % for use with referee option

\end{document}